\documentclass[final,3p,times,twocolumn]{elsarticle}
\usepackage{graphicx}
\usepackage{amsmath}
\usepackage{amssymb}     
\usepackage{hyperref}
\begin{document}
\title{Spin-phonon coupling and dielectric spectroscopy in nano-crystalline Pr$_2$CoMnO$_6$ double perovskite manganite}
\author{Ilyas Noor Bhatti\corref{cor1}}\address{Department of Physics, Jamia Millia Islamia University, New Delhi - 110025, India.}
\ead{inoorbhatti@gmail.com}
\author{Imtiaz Noor Bhatti\corref{cor2}}\address{$^1$Greyc Lab, The French National Center for scientific Research (CNRS),  14000 Caen, France}
\ead{imtiaz-noor.bhatti@unicaen.fr}
\begin{abstract}
 In this paper, we present the spin-phonon coupling and dielectric response of nano-crystalline Pr$_2$CoMnO$_6$ employing Raman and dielectric spectroscopic study. Pr$_2$CoMnO$_6$ is a manganite compound , it undergoes a paramagnetic to ferromagnetic (PM-FM) phase transition around $T_c$ $\sim$172 K. Temperature-dependent Raman scattering experiment  is carried out across $T_c$ to study the spin-phonon behavior in this material. The results from Raman study reveal an obvious softening of the phonon mode involving stretching vibrations of the (Co/Mn)O$_6$ octahedra in ferromagnetic temperature regions, indicating a close correlation between magnetism and lattice in Pr$_2$CoMnO$_6$ and conform the spin phonon coupling. Further, we have carried out detailed study on dielectric response, impedance spectroscopy, electric modulus and AC conductivity of Pr$_2$CoMnO$_6$ ceramics in the temperature range of 20 K - 300 K and frequency range of 1kHz - 5.5MHz. We found Pr$_2$CoMnO$_6$ shows strong frequency dependence with large dispersion and large dielectric constant. It is found that thermally activated relaxation mechanism is involve and material is deviated from Debye's model which is confirmed by Nyquist plot and complex modules behavior.
\end{abstract}

\maketitle

\section{Introduction}
Materials with double perovskite structure are fascinating class of compound shows variety of interesting properties including magnetic,  dielectric, piezoelectric, ferroelectric, optical and magneto-resistive properties.\cite{Roth, Kobayashi, Zhao, Vijaykumar, Kumar, Dutta} Double perovskites have been extensively studied by experimental and theoretical researcher over last two decades. Materials with general formula A$_2$BB$^{\prime}$O$_6$ where A stand for alkaline or alkaline-earth cations, B and B$^{\prime}$ stand for transition elements are known as double perovskite compounds.\cite{Roth}. Material with double perovskite structure have been largely studied for their physical properties, since they have potential to show exotic phenomenon and novel physics. The interesting physical properties make these materials useful for various industrial applications including sensors, non-volatile memories, infra-red detectors, overload protection circuits capacitors, optical switches, relaxors,  radio-frequency filters \cite{Cross, Sahoo, mansuri, Parida}. 
 
Double perovskite A$_2$BB$^{\prime}$O$_6$ where A = La, Pr, Sm, Gd, Ho, Yb, Eu, etc received much attention recently. These compounds show wide range of exciting phenomenon, for instance multiple magnetic phases and intrinsic magneto-dielectric effect in La$_2$CoMnO$_6$.\cite{dass, murthy1} Pr$_2$CoMnO$_6$ shows Griffiths Phase, spin-phonon coupling and enhanced magnetocloric effect. \cite{xin, liu} Magnetocapacitance and magnetoelastic coupling is observed in La$_2$NiMnO$_6$.\cite{nsro, yang} Multiferroicity and magnetism driven ferroelectricity in Y$_2$NiMnO$_6$ was theoretically predicted and experimentally observed.\cite{kumar, su}  Exotic phenomenon like short-range magnetic correlation, giant rotating magnetocloric effect and metamagnetic behavor in (Tb,Nb )$_2$CoMnO$_6$\cite{ranjan, moon} magnetic ordering and magnetoelectric properties are observed in these compounds. \cite{jbl} Further, in thin films of La$_2$CoMnO$_6$ strain-induced perpendicular magnetic anisotropy and spontaneous cationic ordering is observed.\cite{rg, hwang} Magnetism, dielectric properties, order disorder phases ans structure disorder driven properties in double perovskite materials are been focus of studies over the past few decades. Much of the above cited work is on bulk either polly-crystalline or single crystals, however there are not detail study of these compounds at nano-scale. Earlier we have presented our study on structure, magnetic and dielectric response of nano-crystalline (Sm,Gd,Ho,Yb)$_2$CoMnO$_6$ compounds.\cite{ilyas3, ilyas4, ilyas5, ilyas6}. The presence of rare earth elements at A-site in these material tune the magnetic and dielectric properties.\cite{ilyas3, ilyas4, ilyas5, ilyas6, jbl} These materials also show spin/cluster glass behavior and strong spin phonon coupling. Here we focus our study on nano-crystalline Pr$_2$CoMnO$_6$. 

There have been extensive literature available on physical properties of bulk Pr$_2$CoMnO$_6$ however this material is not well studied in nano-crystalline form. In the paper, we present the study of nano-crystalline Pr$_2$CoMnO$_6$ with double perovskite structure. Room temperature X- Ray Photon Spectroscopy is employed to study the charge state of ions present in this compound, spin lattice coupling is studied through temperature dependent Raman Spectroscopy and dielectric response of the nano-crystalline Pr$_2$CoMnO$_6$ is studied in detail.

\section{Experimental details}
Pr$_2$CoMnO$_6$ sample used in this study was prepared by sol-gel method as reported in our earlier work.\cite{ilyas1, ilyas2}. The detail structural study of sample is reported in our work Bhatti etal\cite{ilyas1}. We reported the sample crystallize in single phase and chemically pure. Pr$_2$CoMnO$_6$ adopts orthorhombic crystal structure with Pbnm space group. The lattice parameters a b and c obtained from structural analysis are 5.446(5) Å, 7.695(1) Å and 5.403(4) Å respectively.\cite{ilyas1} In this paper we focus our study on charge distribution, spin-phonon coupling and dielectric repose in Pr$_2$CoMnO$_6$. To explore these properties we have performed room temperature XPS measurements, temperature dependent Raman spectroscopy and dielectric response of Pr$_2$CoMnO$_6$ sample. The XPS measurement were performed with base pressure in the range of $10^{-10}$ mbar using a commercial electron energy analyzer (Omnicron nanotechnology) and a non-monochromatic Al$_{K\alpha}$ X-ray source (h$\nu$ = 1486.6 eV). The XPSpeakfit software was used to analyze the XPS data. The samples used for XPS study are in pallet form where an ion beam sputtering has been done on the samples to expose clean surface before measurements.  Labram-HR800 micro-Raman spectrometer with diode laser having wavelength ($\lambda$) = 473 nm have been used to record the temperature dependent Raman spectra. This spectrometer use grating with 1800 groves/mm and CCD detector with a high resolution of $\sim$ 1 cm$^{-1}$. A THMS600 stage from Linkam UK have been used for temperature variation with stability of $\pm$0.1 K for the low temperature Raman measurements. Dielectric measurements in the frequency range from 1 Hz to 5.5 MHz were performed using a computer controlled dielectric spectrometer.

\section{Result and Discussions}

\subsection{X-ray photo-electron spectroscopy (XPS)}
Physical properties of compound are mostly dependent on oxidation state of ions presents in the compound. Here we have employed the X-ray photo-electron spectroscopic technique to estimated the charge sate of Pr, Co, Mn and O ions in the Pr$_2$CoMnO$_6$. XPS Core level spectra of Co, Mn, Pr and O are shown in Fig. 1.

Co 2$p$ XPS spectra is shown in the Fig. 1a,  we found two peaks corresponding to Co 2$p$$_{3/2}$ and Co 2$p$$_{1/2}$ at 796 eV and 780 eV  respectively. Apart from the main Co peaks satalite peaks are observed at higher energy side from main Co peaks. The peaks Co2$p$ peaks separated by spin-orbital splitting energy of 15 eV and are in agreement with previous literature.\cite{wang, qiu, xia} The peaks locations of Co 2$p$ core level confirm +2 oxidation state of Co cations in Pr$_2$CoMnO$_6$. 

Mn 2$p$ core levels along with peak fitting spectra is present in the Fig. 1b. Here we observed two distinct peaks of Mn $2p$ spectra, one is at 640 eV and 655 eV resulting in splitting of Mn $P$ 2$p$$_{3/2}$ and 2$p$$_{1/2}$  with splitting energy of 15 eV. We also observed two small peaks beside these two peaks which are due to satellite correction and are good agreement with literature.\cite{ida, sachoo, cao}From the peak position we analyzed Mn cations is present in +4 oxidation state.

\begin{figure}[t]
	\centering
		\includegraphics[width=8cm]{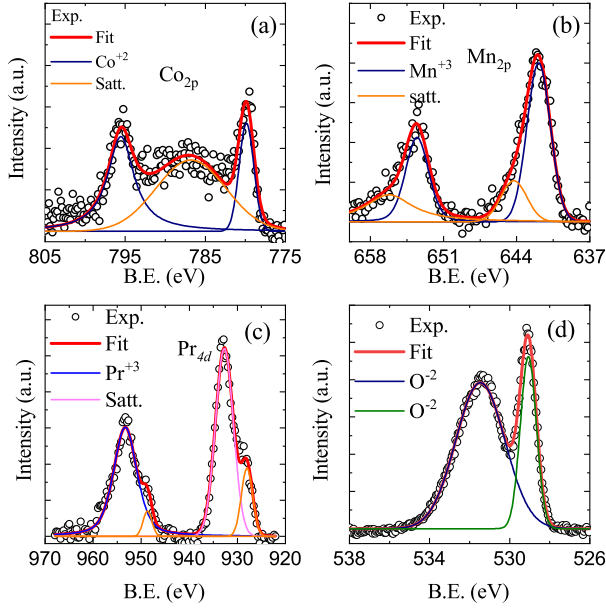}
\caption{(Color online) (a) The XPS core level spectra of Co 2$p$ (b) XPS core level spectra of Mn 2$p$. (c) XPS core level spectra of Pr 4$d$. (d) XPS core level spectra of  O 1$s$.  In the figure the red solid line is the overall envelop of the XPS spectrum and the other colored solid lines are the other respective fitted peaks.}
\label{fig:Fig1}
\end{figure}

Fig. 1c shows the Pr 3d core level XPS spectra. As indicated in figure the two XPS peaks corresponding Pr 3$d$$_{5/2}$ and Pr 3$d$$_{3/2}$ are located at 933.6 eV and 954 eV with spin orbit splitting of 20.4 eV. The peak fitting shows that the Pr cations present in +3 oxidation state in this compound and is good agreement with the literature.\cite{qliu, duan}. In addition, the Pr 3d core level spectra shows a shoulder at lower BE side besides the two peaks of Pr 3$d$$_{5/2}$ and Pr 3$d$$_{3/2}$ as shown in Fig. 1c, theses are called shake-off satellite peaks.

In Fig. 1d XPS spectra of O 1$s$ is presented. There are two peaks at 528 eV and 532 eV are observed. The O 1$s$ peak located at 532 eV corresponding the oxygen anion present in Pr$_2$CoMnO$_6$ whereas the peak at 528 ev is often associated with -OH present as moisture on the sample. From the fitting and peak analysis we found that oxygen is present in -2 oxidation state.

From detailed analysis of XPS spectra of Pr$_2$CoMnO$_6$, results confirm that the charge sate of elements presents in Pr$_2$CoMnO$_6$ are +2, +4, +3 and -2 for Co, Mn, Pr and O respectively. 

\begin{figure*}[t]
	\centering
		\includegraphics[width=15cm]{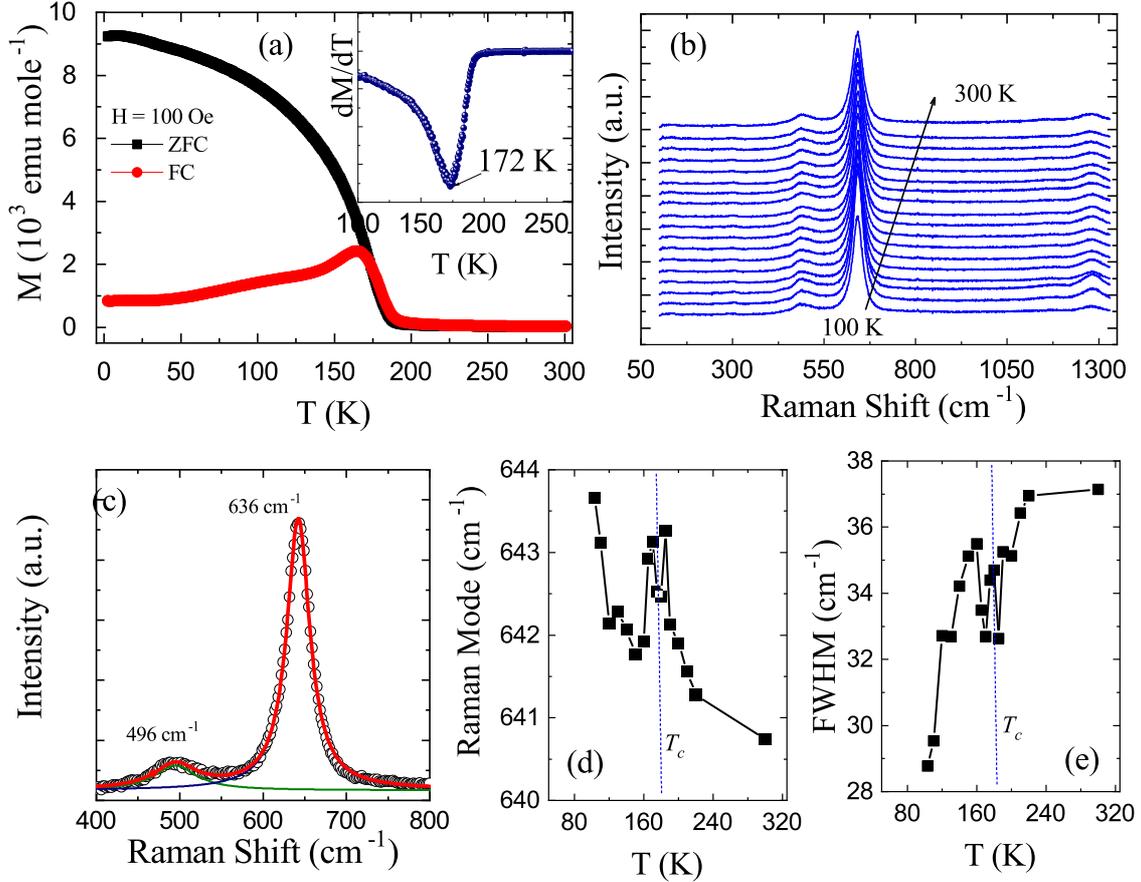}
\caption{(Color online) (a) Temperature dependent magnetization curve M(T) for Pr$_2$CoMnO$_6$ measured at 100 Oe applied magnetic field. Inset is the $dM/dT$ vs $T$ where point of infliction gives the transition temperature. (b) Raman spectra of Pr$_2$CoMnO$_6$ measured at different temperatures. (c) Shows the line shape and its Lorentzian fitting of A$_{1g}$ and B$_{2g}$ Raman modes at 494 and 644 cm$^{-1}$ respectively. (c) Temperature Variation of (d) Raman shift (e) FWHM for Raman mode at 640 cm$^{-1}$ corresponding to stretching of Co/MnO$_6$ for Pr$_2$CoMnO$_6$. The solid line is fitting due to Eq. 1}
	\label{fig:Fig2}
\end{figure*}

\subsection{Temperature Dependent Raman Study}
Fig. 2a shows M(T) curve of Pr$_2$CoMnO$_6$ under applied magnetic field of 100 Oe, we observed a paramagnetic to ferromagnetic phase transition around 172 K in this material. The phase transition is due to Mn$^{4+}$ and Co$^{2+}$ ordering, there is no anomaly observed at lower temperature which is expected in site disordered double perovskite due to anti-ferromagnetic interaction among different oxidation states of Mn cations. This goes well with our XPS results that confirms only Mn$^{4+}$ and Co$^{2+}$ oxidation states present in the sample. The detail study of magnetic properties of same sample is reported in our earlier work.\cite{ilyas1, ilyas2} To study the spin-phonon coupling in this material we have measured the Raman spectra over a wide range of temperature across magnetic transition. Fig. 2b shows the temperature dependent spectra of Pr$_2$CoMnO$_6$. We have taken the data across magnetic phase transition 80 K to 300 K with an interval 5 K close to the phase and away from phase transition is 10 K. From the figure we observe that there are two mode present in the spectra and an over tone. One mode is prominent which is located at 644 Cm$^{-1}$ and the weak mode present in 495 Cm$^{-1}$ as shown. W have also observed an overtone at 1296 Cm$^{-1}$ in full temperature range. The prominent Raman modes present at  695 cm$^{-1}$ is corresponding to B$_{2g}$ stretching mode and weak mode present at 495 Cm$^{-1}$  is  A$_{1g}$ breathing mode. These two weak and Prominent Raman modes are due to stretching, bending and rotation of (Co/Mn)O$_6$ octhadera in the compound Pr$_2$CoMnO$_6$. The strong mode present in the compound is due to symmetric stretching of the (Co/Mn)O$_6$ octahedra and weak mode is due to a mixed type vibration of antisymmetric stretching and bending of the (Co/Mn)O$_6$. The fitting of Raman mode at 695 cm$^{-1}$is done using Lorentzian fit as shown in Fig. 2c. Full with at half maxima (FWHM) and mode position is obtained from the fitting parameters and plotted as function of temperature in Fig. 2d and 2e. It is obvious from the Fig. 2d and 2e that there is deviation in mode position and FWHM around $T_c$ such deviation is often resulted from spin phonon coupling. To further understanding the spin phonon coupling in the compound we a have fitted the temperature variation of mode position and FWHM following anharmonic decay model:\cite{harish}
\begin{eqnarray}
\omega(T) = \omega{_0} - A\left[1 + \frac{2}{exp\left(\frac{\hbar\omega_0}{2k_BT}\right) - 1}\right]
\end{eqnarray}

\begin{eqnarray}
\Gamma(T) = \Gamma{_0} - B\left[1 + \frac{2}{exp\left(\frac{\hbar\omega_0}{2k_BT}\right) - 1}\right]
\end{eqnarray}

where $\omega_0$ is intrinsic frequency and $\Gamma_0$ is line width of the optical mode, A and B are the anharmonic coefficients. $\omega(T)$  and $\Gamma(T)$ describes expected temperature dependence of a phonon mode frequency and line width  due to anharmonic phonon-phonon scattering. 
In Fig. 2d Raman mode as function of temperature shows a deviation around magnetic phase transition $T_c$ $\sim$170 K. We further plot the FWHM vs temperature her also a deviation in the from anharmonic behavior around $T_c$. Such behavior indicated the presence of spin phonon coupling in the material which is in good agreement with the literature \cite{sandi, grana, lave}.

\begin{figure*}[ht]
	\centering
		\includegraphics[width=15cm]{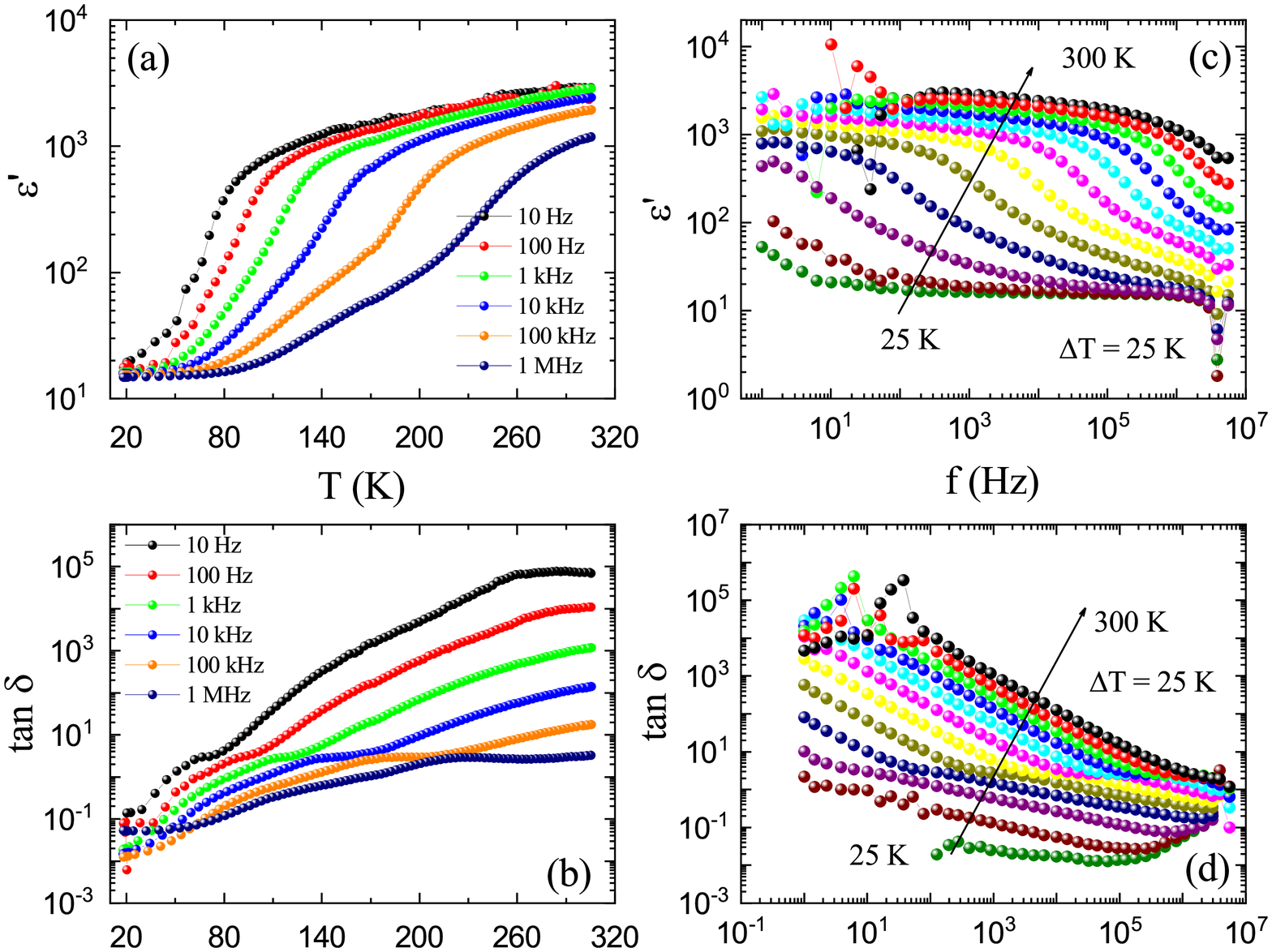}
\caption{(Color online) Temperature dependent (a) real part of complex dielectric permittivity $\epsilon^{\prime}$. (b) dielectric loss tan $\delta$ measure for Pr$_2$CoMnO$_6$ in the temperature range of 20 K to 300 K at various frequencies. (c) Frequency dependent $\epsilon^{\prime}$ measured at different temperatures. (d) Frequency dependent $\epsilon^{\prime \prime}$ measured at different temperatures.}
	\label{fig:Fig3}
\end{figure*}
\begin{figure} 
	\centering
		\includegraphics[width=8cm]{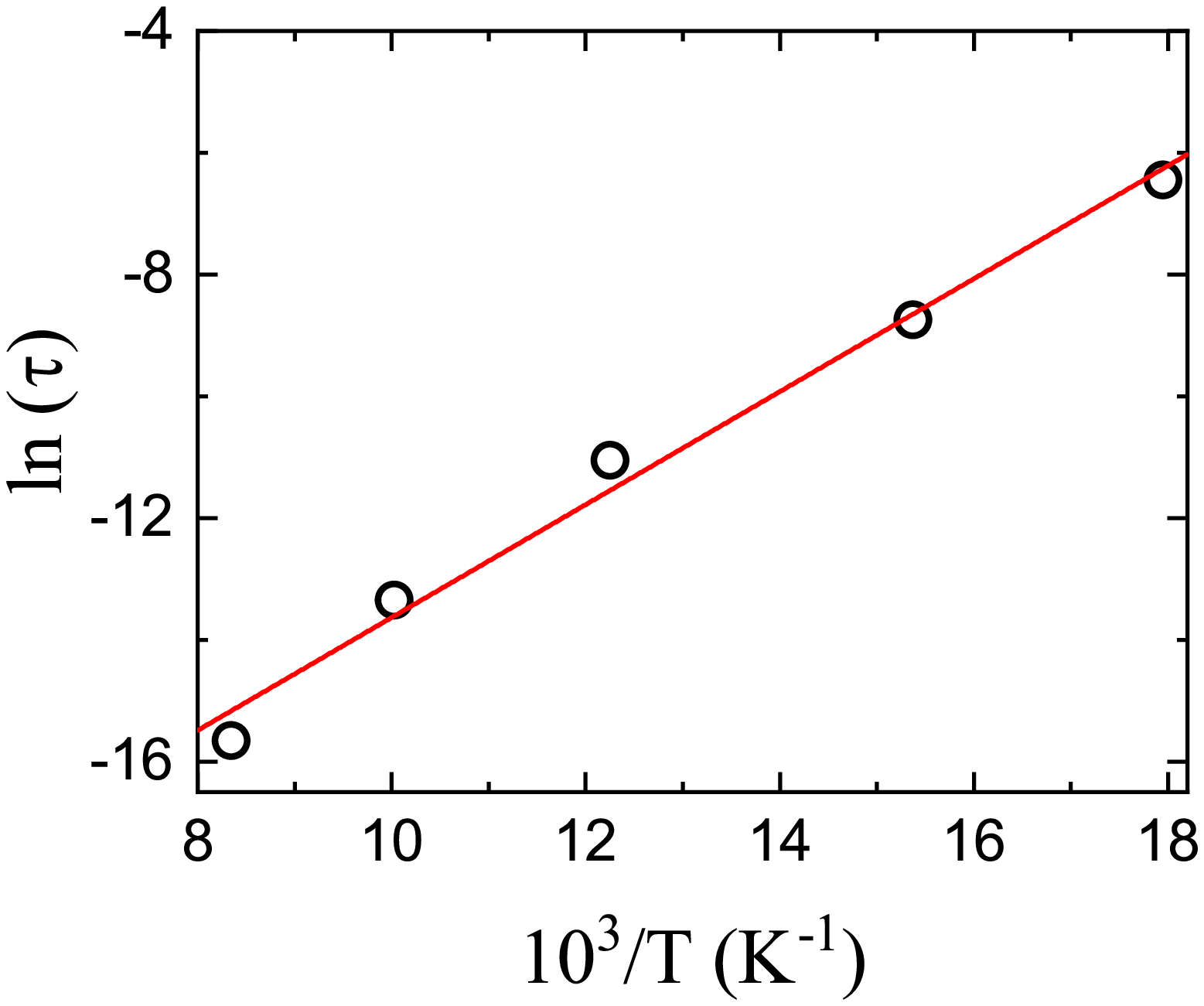}
\caption{(Color online) Variation of relaxation time against normalized temperature i.e ln $\tau$ vs 1000/T obtained from tangent loss plot.}
	\label{fig:Fig4}
\end{figure}

\subsection{Dielectric study}
In this section we focus our study on temperature and frequency dependent dielectric response of Pr$_2$CoMnO$_6$. Fig. 3a shows the temperature dependent real part of dielectric permitivity $\epsilon^{\prime}$ measured at different frequencies between 10 Hz to 1 MHz. From the figure it is observed that with increasing frequency the $\epsilon^{\prime}$ decreases, further with increasing temperature dielectric permitivity increases at all frequencies. At sufficient low low temperature the $\epsilon^{\prime}$ curves measured at different frequencies are merge into one curve and at higher temperature the curves are closing to each other.
Fig. 3b shows the dielectric loss (tan $\delta$) as a function of temperature measured at different frequencies. From the figure it is evident that with increasing temperature and decreasing frequency tan $\delta$ increases. There is relaxation process indicated by appearance of broad peak in tan $\delta$ and the peaks shifts to higher temperature with increasing frequencies this confirms a thermally activated relaxation process involved in this material.  Fig. 3c and 3d plots shows frequency dependent dielectric permittivity $\epsilon^{\prime}$ and dielectric loss $\delta$ from 20K to 300K temperature range we measured $\epsilon^{\prime}$ and $\delta$ at different frequency from 10Hz to 1MHz for Pr$_2$CoMnO$_6$ material. In the figures 3c. and 3d we observed with increasing temperature the dielectric permittivity $\epsilon^{\prime}$ and $\epsilon^{\prime\prime}$ increase, where as with increasing frequency it shows decreasing trend. The higher value of $\epsilon^{\prime}$ at low frequency is attributed to the accumulation of the charges at grain boundaries. There is strong dispersion in mid frequency regime. In tan $\delta$ vs $f$ response we observed peaks and these peaks shifts to higher frequency with increasing temperature. In dielectric loss tan $\delta$ it is clearly seen that there is a broad hump at low temperatures which is feature of relaxor phenomenon. This clearly indicates that the relaxation peaks in the dielectric loss is due to thermally activated mechanisms. The resonance condition defined as $\omega_p$$\tau_p$ = 1 where $\omega$ = 2$\pi$$f$ is defined as resonance frequency. The relaxation mechanisms and its origin can be analyzed by fitting the peaks in tan $\delta$ with the Arrhenius law given as follow:
\begin{eqnarray}
 \tau_{tan \delta} = \tau_0 exp\left(\frac{-E_{\alpha}}{k_B T}\right)
\end{eqnarray}
where, $\tau_{tan \delta} = \frac{1}{2 \pi f_{tan \delta}}$, $T$ is the temperature where peak occurs in tangent loss curve at a particular frequency $f_{tan \delta}$, $\tau_0$ and $E_{\alpha}$ are characteristic relaxation temperature and activation energy respectively and, $k_B$ is the Boltzmann constant. Fig. 4 shows the plot of dielectric loss peaks as a function of absolute temperature. From the fitting parameters of the data using Eq. 3 we have calculated activation energy $E_{\alpha}$ = 0.68 eV. 

\subsection{Impedance spectroscopy}
Impedance spectroscopy is vital and informative technique to understand and distinguish the contributions to the electric and dielectric properties from  grain, grain boundaries and electrode-sample contact interface. Various involve in relaxation mechanisms can be identified by plotting impedance in a complex plan at various temperature. The complex impedance describe by the equation:\cite{fang}

\begin{figure}[ht]
	\centering
		\includegraphics[width=8cm]{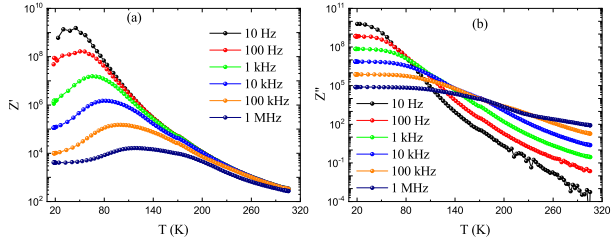}
\caption{(Color online) (a) Temperature dependent real part of impedance $Z^{\prime}$ measure at different frequencies. (b) Temperature dependent real part of impedance $Z^{\prime \prime}$ measure at different frequencies.}
	\label{fig:Fig5}
\end{figure}

\begin{eqnarray}
Z^* = Z^\prime + jZ^{\prime\prime}, Z^\prime = \frac{R}{1 + (\omega\tau)^2}, Z^{\prime\prime} = \frac{\omega R \tau}{1 + (\omega\tau)^2}
\end{eqnarray}
where Z$^*$ is complex impedance, Z$^\prime$ and Z$^{\prime\prime}$ are real and imaginary parts of impedance respectively. R is resistance, $\omega$ is angular frequency and $\tau$ is relaxation time.
Fig. 5a and Fig 5b shows the real  and imaginary part of complex impedance (Z$^{\prime}$) and (Z$^{\prime\prime}$) plotted as a function of temperature in the frequency range 1 Hz to 1 MHz. It is evident from the Fig. 5a that Z$^{\prime}$ decreases with increasing temperature. At low temperature Z$^{\prime}$ gradually decreases with increasing frequency, however at temperatures above 150 K, Z$^{\prime}$ curves merge into one curve. Fig. 5b represent the temperature dependent Z$^{\prime\prime}$ plot measured at different selective frequencies. The curve show dispersion at low and high temperature. The Z$^{\prime\prime}$ shows a decreasing trend with increasing temperature which is suggesting enhance conduction at high temperature due to thermally activated charge carriers in this material. Further, in Fig 6a and 6b we have plotted Z$^{\prime}$ and Z$^{\prime\prime}$ as function of frequency measured at different temperatures in between 20 K to 300 K. The figures shows a dispersion a low and mid frequency region where at high frequency we see the curves are closely approaching each other.  At low frequency the Z$^{\prime}$ remains frequency independent and this pleatu region moves to higher frequency with increasing temperature. Which is due to the release of accumulated space charges at high temperatures hence contribute to the enhancement of conduction in this material at high temperature. Fig. 6b shows the imaginary part of impedance ($Z^{\prime\prime}$) as a function of frequency. We observed that the $Z^{\prime\prime}$ curves show a peak value corresponding to a particular frequency. This feature is seen in for all the $Z^{\prime\prime}$ curves measured at different temperatures and observed this peak moves towards higher frequency with increasing temperature. The peak shift to higher frequency with increasing temperature suggests that the relaxation time constant decreases with increasing temperature. The peak frequency is used to calculate the relaxation time and we found the relaxation time $\tau$ follows Arrhenius behavior. The relaxation time vs absolute temperature is fitted and fitted parameter are used to calculate the activation energy E$_a$ = 0.76 eV. This value is in agreement with values obtained from dielectric loss data. 
\setlength{\tabcolsep}{9pt}
{
\begin{table*}
\begin{tabular}{c c c c c c c c}
\hline
\hline
T (K) &C1 &C2 &R1 &R2 &R3 &P1 &n1\\
\hline
300 &7.8844E-11 &1.9723E-12 &1.9816E-14 &261.6 &114.69\\
275 &0.0005 &4.1637E-11 &9.8886E-13 &460.23 &413.69 &1.9671E-09 &0.93975\\
250	&0.00049335	&2.1765E-11	&3.5049E-12	&951.18	&1000.8	&1.6894E-09 &0.93038\\
225	&6.709E-05	&4.4382E-09	&2.2667E-12	&2693.4	&2281.4	&1.9929E-09 &0.90414\\
200	&9.7686E-12	&5.4005E-10	&2.2302E-11	&9672.1	&5164.9	&8.699E-08 &0.59478\\
175	&6.7766E-10	&0.0005	&4.808E-05	&33601	&21370	&1.8604E-09 &0.66258\\
150	&7.003E-10	&0.00048051	&5.518E-11	&91445	&1.6964E05	&1.2125E-09 &0.65886\\
125	&2.4663E-11	&4.3655E-10	&8.3179E-11	&7.7148E05	&1.0098E06	&2.2123E-09 &0.58978\\
100	&1.0363E-10	&5.3067E-10	&8.3179E-10	&7.7148E05	&1.0098E06	&2.2123E-09 &0.58978\\
\hline
\hline
\end{tabular}
\caption{ Values obtained for the components of equivalent circuit from Nyquist plot fitting for Pr$_2$CoMnO$_6$.}
\label{tab:table 1}
\end{table*}
} 
\begin{figure} 
	\centering
		\includegraphics[width=8cm]{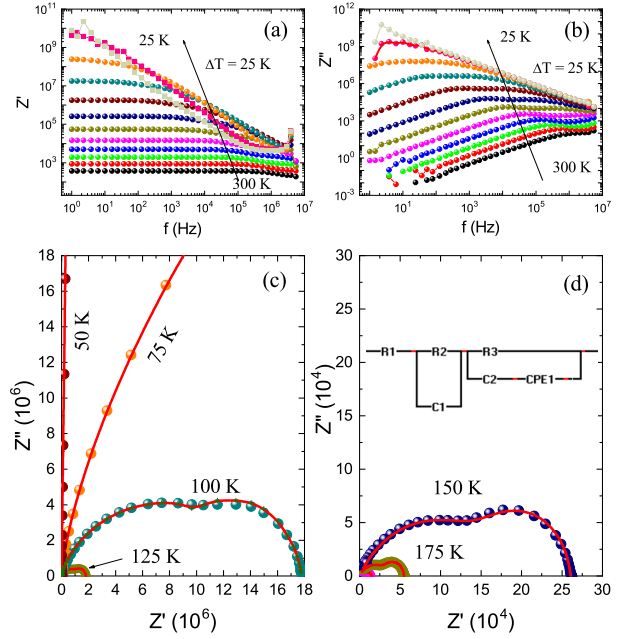}
\caption{(Color online) (a) Frequency dependent real part of impedance Z$^{\prime}$ measure at different temperatures. (b) Frequency dependent imaginary part of impedance Z$^{\prime\prime}$ measure at various temperatures. (c), (d) real Z$^{\prime}$ and imaginary part Z$^{\prime\prime}$ plotted in terms of Nyquist plot Z$^{\prime}$ vs Z$^{\prime\prime}$ and inside (d) Equivalent circuit of Nyquist plot for Pr$_2$CoMnO$_6$.}
	\label{fig:Fig6} 
\end{figure}

\begin{figure}[t] 
	\centering
\includegraphics[width=8cm]{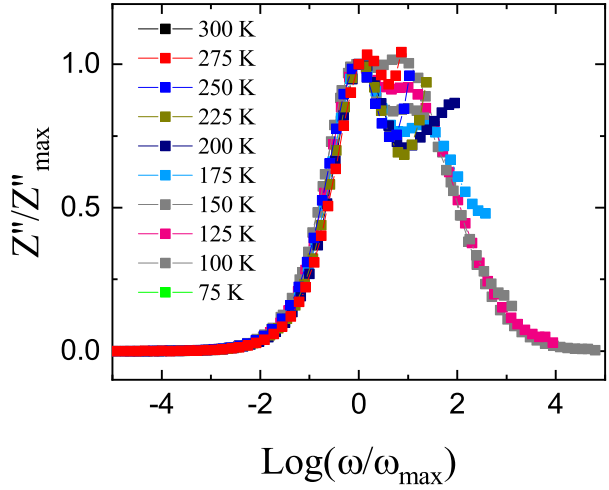}
\caption{(Color online) Scaling behavior of imaginary part of impedance $Z^{\prime\prime}$ i.e. i.e. Z$^{\prime\prime}$/Z$^{\prime\prime}_{Max}$ vs ln$\omega$/$\omega_{Max}$ for Pr$_2$CoMnO$_6$.}
	\label{fig:Fig7}
\end{figure}
\subsection{Nyquist plot}
The complex impedance spectra i.e. Z$^{\prime\prime}$ vs Z$^\prime$ (Nyquist plot) for Pr$_2$CoMnO$_6$ at representative temperatures is shown in Fig. 6c and 6d. Nyquist plot signifies the variation of Z$^{\prime\prime}$ with Z$^\prime$ in the measured  frequency range at different temperature of 20 K–300 K. The Nyquist plot above 100 K  shows two semicircle whereas below this temperature incomplete circles are observed which is due to highly insulating nature of material.  The radii of Nyquist plot shows a decreasing trend with increasing temperature. The two semicircle in the Nyquist plots suggest the contribution from both intrinsic (grain interior) bulk properties as well as grain boundaries in the material. The asymmetric nature of semicircles and origin lying below x-axis which suggests non-Debye type of relaxation mechanism in this material. It also manifests that there is a distribution of relaxation time instead of a single relaxation time in the material. The equivalent circuit for the Nyquist plot represented inside Figs. 6d. We found Equivalent circuit consists of  series and parallel combination of resistor, constant phase change and capacitor. We have fitted the Nyquist plot with equivalent circuit and obtained values of components are tabulated in the table-1. Fig. 7 shows the scaling behavior of imaginary part of impedance we observed that the impedance curve measured at different temperatures erg into one master curve. This scaling behavior suggest same relaxation mechanism in full range of temperature.

	\begin{figure} [th]
	\centering
		\includegraphics[width=8cm]{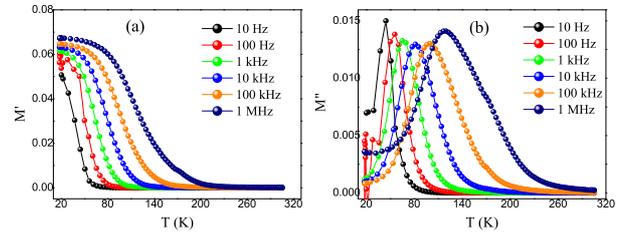}
\caption{(Color online) (a) Temperature variation of real part of electrical modulus i. e. M$^\prime$ vs T. (b) Temperature variation of imaginary part of electrical modulus i.e. M$^{\prime\prime}$ vs T.}
	\label{fig:Fig8}
\end{figure}
\begin{figure} [h]
	\centering
		\includegraphics[width=8cm]{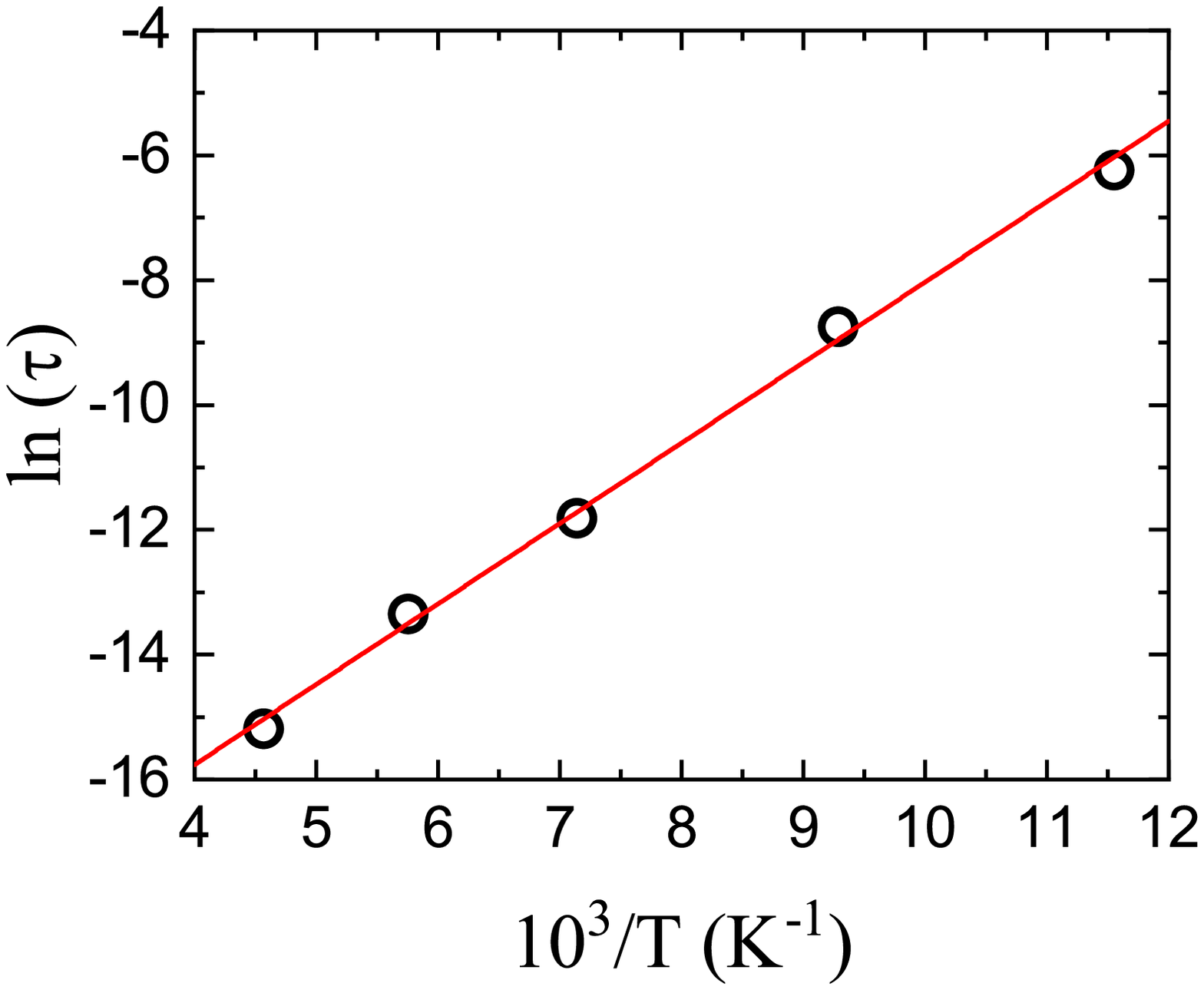}
\caption{(Color online) Variation of relaxation time against normalized temperature i.e ln $\tau$ vs 1000/T obtained from M$^{\prime\prime}$ plot. }
	\label{fig:Fig9}
	\end{figure}
	
	\begin{figure} [th]
	\centering
		\includegraphics[width=8cm]{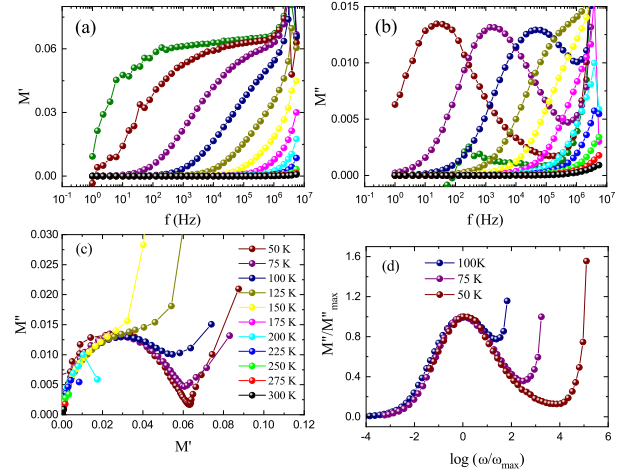}
\caption{(Color online)  Variation of electrical modulus with frequency (a) real part i.e. M$^\prime$ vs f. (b) imaginary part i.e. M$^{\prime\prime}$ vs f. (c) shows the nyquist plot of electrical modulus i.e. M$^\prime$ vs M$^{\prime\prime}$ (d) Scaling behaviour of M$^\prime\prime$ at various temperatures i.e. M$^{\prime\prime}$/M$^{\prime\prime}_{Max}$ vs ln$\omega$/$\omega_{Max}$ for Pr$_2$CoMnO$_6$.}
	\label{fig:Fig10}
\end{figure}
	
\subsection{Modulus spectroscopy}
The study electric modulus of a material is essential to observe the electrode effect, interface polarization, conductivity, grain boundary conduction effect and relaxation behavior under frequency and temperature constrain. Here we present temperature and frequency dependant modulus analysis of Pr$_2$CoMnO$_6$. Fig. 8a and 8b represent temperature dependent real (M$^{\prime}$) and imaginary (M$^{\prime\prime}$) parts of electric modules for Pr$_2$CoMnO$_6$ measure in temperature range of 20 K to 300 K at different frequencies between  10 Hz and 1 MHz.  In Fig. 8a it found that (M$^{\prime}$) reaches a constant value at high temperature where all curves merge for Pr$_2$CoMnO$_6$.  M$^\prime$ increases with decreasing temperature at low temperatures all the curves of M$^\prime$ measured at different frequencies are approaching each other.  There is large dispersion in M$^\prime$ at intermediate temperatures which is due to short-range charge carriers involve in relaxation process also evident in frequency dependent M$^{\prime}$ curves. Fig. 8b shows temperature variation of $M^{\prime\prime}$ at selected frequency. $M^{\prime\prime}$ plot reveals relaxation phenomena in the material as the curve attain a peak value and this peak moves to higher temperature with increasing frequency. This behavior suggests that hopping of charge carriers is predominantly thermally activated. We observed asymmetric broadening of the peak that is suggesting relaxation process with different time constants, this feature again confirms the non-Debye type relaxation in this material. In Fig. 9 we have plotted the $\tau$ vs absolute temperature $T$ calculated from the peak frequency from Fig. 8b following the relation $\tau$ = 1/2 $\pi$ f. We observed that the relaxation time as a function of absolute temperature (1000/T) obeys Arrhenius law. From the fitting parameters we found the activation energy  E$_\alpha$ = 0.73 ev, which is comparable with the value obtained from impedance data. 

Fig. 10a and 10b shows the real and imaginary part of electric modulus as a function of frequency at different temperatures. Fig. 10a shows strong dispersion in mid frequency range and at low and high frequencies the curves merge in to single curve. This dispersion is attributed to the presence of short-range charge carriers that are involved in the relaxation process. In Fig. 10b it is evident that the  $M^{\prime\prime}$ attain a peak values which shifts to higher frequency with increasing temperature. This shifting of peak to higher frequencies suggest that the thermally activated relaxation mechanism is active in this material.  In Fig. 10c we have presented the complex modules plot i.e $M^{\prime\prime}$ vs $M^{\prime}$ at different temperatures. The shape of complex modules plot shows two semicircles one complete and tail of second semicircle suggest the contribution from grains and grain boundaries effect. The asymmetric semicircles further confirm the non-Debye's type of relaxation in this material.

Scaling behavior study of electric modulus is a vital tool to understand the relaxation dynamics in dielectric materials. The temperature and particle-size-dependency of relaxation dynamics can be evaluated by scaling analysis.  The scaling behaviour of  M$^{\prime\prime}$ at various temperatures i.e. M$^{\prime\prime}$/M$^{\prime\prime}_{Max}$ vs ln$\omega$/$\omega_{Max}$ for Pr$_2$CoMnO$_6$ is plotted in Fig. 10d. The M$^{\prime\prime}$ curves measure at different temperatures merge into single master curve. This indicates that the dynamics processes are temperature independent i.e. material shows same relaxation mechanism at all temperatures.  Inset of Fig. 10d shows the comparative plot of normalized imaginary parts of modulus (M$^{\prime\prime}$/M$^{\prime\prime}_{Max}$) and impedance Z$^{\prime\prime}$/Z$^{\prime\prime}_{Max}$ as a function of frequency at 100 K for Pr$_2$CoMnO$_6$. The peak frequency of electric modulus and impedance are slightly separated, this feature of comparative plots confirm that the material have contribution from both long-range and localized relaxation process. The equivalent circuit involve constant phase element for  the fitting of Nyquist plot is again justified by this peak separation in Z$^{\prime\prime}$ and M$^{\prime\prime}$ plot.

\begin{figure}
\centering
	\includegraphics[width=8cm]{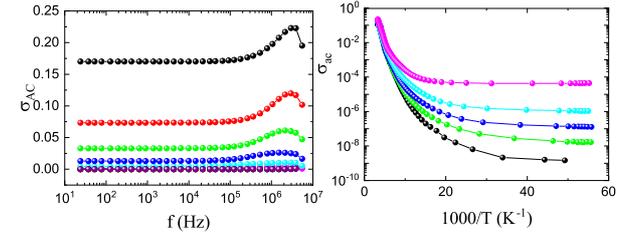}
\caption{(Color online) (a) Frequency dependence ac conductivity i.e. $\sigma_{ac}$ vs $f$ at selective temperatures between. The solid lines are fitting due to Eq. 5. (b) The variation of $\sigma_{ac}$ with absolute temperature (1000/T) at different frequencies are shown for Pr$_2$CoMnO$_6$. The solid lines are fitting due to Eq. 6.}
	\label{fig:Fig11}
\end{figure}

\subsection{AC conductivity} 
The temperature and frequency dependent ac conductivity ($\sigma_{ac}$) of Pr$_2$CoMnO$_6$ is plotted in Fig. 11. The ac conductivity is calculated from dielectric data using the relation $\sigma_{ac} = \epsilon_0 \omega \epsilon^{\prime\prime}$.\cite{sing} The frequency dependent $\sigma_{ac}$ at different temperature is plotted in Fig. 11a. As evident from the figure at low frequencies the conductivity is independent of frequency and dc conductivity ($\sigma_{dc}$) is accounts for conduction process at low frequencies. A plateau region is observed for all temperatures at low frequencies which notably extend to high frequency with increasing temperature. The polarization effect at the dielectric interface i.e grain boundaries and electrode is account for variation of conductivity at low frequencies. Further, we observed that at higher frequencies the conductivity increases with further increase in frequency by reducing the charge accumulation. The frequency independent region also suggests that the hopping charge carriers are absent at low frequencies. The ac conductivity at high frequency in this case obey Jonscher's Universal Power Law given as follow:\cite{Jonscher}
\begin{eqnarray}
\sigma_{ac} = \sigma_{dc} + A\omega^n
\end{eqnarray}
when A is a temperature dependent constant, $\omega$ = 2$\pi$ $f$ and $n$ is the power law exponent and its value lies between 0 and 1. $n$ = 1 for pure Debye's model and $<$ for interacting system particularly non-Debye's in nature. The frequency dependent $\sigma_{ac}$ is fitted with Eq. 5 shown as solid lines in Fig. 11a. The conductivity data is fitted well  in full frequency range for temperature $>$200 K below this temperature data does not obey power law. From the fitting parameters we obtained the values of exponent factor $n$, A and $\sigma_{dc}$ for all fitted curves and tabulated in Table-2. The observed $n$ is less than 1 which confirms the material behaves as non-Debye's system. Further, general trend of increasing A and decreasing $n$ values are observed with decreasing temperature. 

To further understand the conduction mechanism we have plotted the variation of ac conductivity with absolute temperature i.e.  $\sigma_{ac}$ vs 10$^3$/T at selective frequencies ranging from 10 Hz to 1 MHz show in Fig. 11b. It is evident from the figure that the observed ac conductivity increases with increasing frequency and temperature. At low temperature the $\sigma_{ac}$ shows strong dispersion. However, at high temperature above 200 K there is sharp increase in the value and the $\sigma_{ac}$ curves at different frequencies tend to merge in to one curve. It seems the $\sigma_{ac}$ shows strong temperature dependency at high temperature and independent of frequency. The thermally activated charges carriers exchange interaction leads to the enhanced $\sigma_{ac}$ at high temperature. The activation energy is calculated by fitting the temperature dependent $\sigma_{ac}$ using Arrhenius relation:
\begin{eqnarray}
\sigma_{ac} = \sigma_{0}exp\left(\frac{-E_\alpha}{k_BT}\right)
\end{eqnarray}
where $\sigma_{0}$ is pre-exponent factor, $k_B$ is Boltzmann constant and $E_\alpha$ is the activation energy.
In Fig 11b the solid lines ar fitting due to Eq. 6. It is found that the conductivity data is well fitted at high temperatures for all frequency. At low temperature the AC conductivity is independent of temperature and form a steady region but does depend on frequency. From the fitting parameters we found the activation energy have inverse relation with frequency i.e. E$_\alpha$ increases with decreases frequency. The calculated values of activation energy are 0.58, 0.63, 0.68, 0.72, 0.75 eV for 1 M, 10 k, 1 k, 100, 10 Hz frequencies respectively.

\setlength{\tabcolsep}{12pt}
{
\begin{table}
\begin{tabular}{c c c c}
\hline
\hline
T (K) &$\sigma_{dc}$ ($\Omega^{-1}$ cm$^{-1}$) &A &n\\
\hline
300 &0.1694 &5.8132E-7 &0.77\\
275 &0.0726 &1.8702E-6 &0.68\\
250 &0.0323 &2.9684E-6 &0.63\\
225 &0.01231 &7.3472E-6 &0.52\\
200 &0.0036 &7.3472E-4 &0.27\\
\hline
\hline
\end{tabular}
\caption{ Tabulated the value of dc conductivity ($\sigma_{dc}$), coefficient (A) and exponent (n) is obtained by fitting ac conductivity ($\sigma_{ac}$) data for Pr$_2$CoMnO$_6$.}
\label{tab:table 2}
\end{table}
} 
It is worth notable that the results from different techniques agreeable from XPS study we found Mn and Co are present in +4 and +2 charge state respectively. The absence of mixed valency of Mn cations reduces the chances of anti-site disorder and hence low temperature additional antiferromagnetic (AF) ordering. The temperature dependent magnetic study reveals no AF magnetic transition due to mixed valencies. Since we observed PM to FM magnetic phase transition around 172 K, Raman scattering data show deviation from anhormonicity at $T_c$ which confirms a lattice-spin coupling. Dielectric study shows that this material is non-Debey's in nature.

\section{Conclusion}
We have studied the 3$d$ based nano-crystalline double perovskite Pr$_2$CoMnO$_6$ employing X-ray photon spectroscopy (XPS) for charge state confirmation, Raman spectroscopy for spin phonon coupling and dielectric response. From X-ray photon spectroscopy (XPS) detailed analysis we found compound elements in Pr$_2$CoMnO$_6$ have charge state  +2, +4, and +3 and -2 for Co, Mn, and Pr, O respectively. Pr$_2$CoMnO$_6$ shows a PM to FM phase transition at T$_c$ $\sim$172 K. Raman spectroscopy gives evidence of strong spin-phonon coupling present in the material which is reflected in deviation of FWHM and mode position  from harmonic behavior. Further dielectric response is studied in detail we observed that the material have high dielectric constant at at room temperature and shows strong dispersion in mid frequency range. The dielectric loss show a relaxation process due to grains, which is thermally activated in nature. The impedance spectroscopy and electrical modulus study further confirms that the relaxation process is thermally activated in nature where the relaxation time follows Arrhenius behavior. Impedance spectroscope and electric modulus study reveal that system is non-Debye model in nature and Nyquist plot shows contribution from intrinsic i.e. from inside grains as well as grain boundaries. AC conductivity have been studied with both frequency and temperature dependence. We found that the activation energy decreases with increasing frequency where as exponent factor $n$ is less then 1 which signifies non-Debye's nature of Pr$_2$CoMnO$_6$.  

\section{Declaration of competing interest}
We would like to declare that we do not have any commercial or associative interest that represents a conflict of interest in connection with the manuscript entitled “Spin-phonon coupling and dielectric spectroscopy in Pr$_2$CoMnO$_6$ double perovskite manganite”.
\section{CRediT authorship contribution statement}
Ilyas Noor Bhatti: Writing - original draft, Data curation, formal
analysis. Imtiaz Noor Bhatti: Formal analysis, Investigation,, Methodology, supervision, Writing-review and editing.
\section{Acknowledgment}
We acknowledge Malaviya National Institute of Technology Jaipur (MNIT Jaipur), India for XPS data and AIRF (JNU) for magnetic measurement. We acknowledge UGC-DAE-Consortium Indore and Dr. V. G. Sathe for Raman data. We also acknowledge Dr. A. K. Pramanik for dielectric measurement and UPEA-II funding for LCR meter. We also aknowledge Dr. Rabindra Nath Mahato for sample synthesis. Author Ilyas Noor Bhatti acknowledge University Grants Commission, India for financial support. 

\end{document}